\documentclass[aps,prb,showpacs]{revtex4}
\usepackage{amsmath,amssymb}
\usepackage{graphicx}
\usepackage{dcolumn}
\usepackage{bm}
\usepackage{color}
\newcommand{\mbb}{\mathbb}
\newcommand{\pa}{\parallel}
\newcommand{\tet}{\texttt}
\newcommand{\mc}{\mathcal}
\makeatletter
\newcommand*{\rom}[1]{\expandafter\@slowromancap\romannumeral #1@}
\makeatother

\begin{document}

\title{Controlling plasmon modes and damping in buckled two-dimensional material open systems}

\author{Andrii Iurov}
\email{aiurov@unm.edu}
\affiliation{
Center for High Technology Materials, University of New Mexico,
1313 Goddard SE, Albuquerque, NM, 87106, USA
}
\author{Godfrey Gumbs}
\affiliation{Department of Physics and Astronomy, Hunter College of the City
University of New York, 695 Park Avenue, New York, NY 10065, USA}
\affiliation{Donostia International Physics Center (DIPC),
P de Manuel Lardizabal, 4, 20018 San Sebastian, Basque Country, Spain}
\author{Danhong Huang}
\affiliation{Air Force Research Laboratory, Space Vehicles Directorate,
Kirtland Air Force Base, NM 87117, USA}
\affiliation{Center for High Technology Materials, University of New Mexico,
1313 Goddard SE, Albuquerque, NM, 87106, USA}
\author{Liubov Zhemchuzhna}
\affiliation{Department of Physics and Astronomy, Hunter College of the City University of New York,
695 Park Avenue, New York, NY 10065, USA}

\date{\today}

\begin{abstract}
Full ranges of both hybrid plasmon-mode dispersions and their damping are studied systematically by our recently developed mean-field theory in open systems involving a conducting substrate and 
a two-dimensional (2D) material 
with a buckled honeycomb lattice, such as silicene, germanene, and a group \rom{4} dichalcogenide as well.
In this hybrid system, the single plasmon mode for a free-standing 2D layer is split into one acoustic-like and one optical-like mode, leading to a dramatic change in the damping of plasmon modes.  
In comparison with gapped graphene, critical features associated with plasmon modes and damping in silicene and molybdenum disulfide are found 
with various spin-orbit and lattice asymmetry energy bandgaps, doping types and levels, and coupling strengths between 2D materials and the conducting substrate. 
The obtained damping dependence on both spin and valley degrees of freedom is expected to facilitate measuring the open-system dielectric property and the spin-orbit coupling strength of individual 2D materials. 
The unique linear dispersion of the acoustic-like plasmon mode introduces additional damping from the intraband particle-hole modes which is absent for a free-standing 2D material layer,  
and the use of molybdenum disulfide with a large bandgap simultaneously suppresses the strong damping from the interband particle-hole modes.
\end{abstract}


\maketitle

\section{Introduction}
\label{sect1}

Plasmons, or self-sustained electron density oscillations, represent a broad interest and have
become an important subject for both traditional and recently discovered two-dimensional (2D) materials.\,\cite{fstern,
SDS07,wunsch,pavlo1,stauber1,stauber2,SDSft,mos2} Due to the possibility of fabricating complex stacking layered structures, 
low-dimensional materials have become very attractive for novel quantum-electronic devices. 
Discovery of graphene, successfully fabricated in 2004, has initiated a number of new
transport and optical studies\,\cite{gr1,gr2,gr3} due to its unusual 
electronic properties originating from the relativistic linear dispersion of its energy bands. 
In particular, graphene plasmonics has quickly become one of the actively-pursued research focuses. 
As an example, novel optical devices in a wide-frequency range showed 
significant improvement in all the crucial device characteristics (see Ref.\,[\onlinecite{poliNS}] and the references
therein). At high energies, the $\pi$-bond plasmons in graphene demonstrate 
both anisotropy and splitting.\,\cite{desp} Moreover, a plasmonic nanoarray coupled to a single graphene sheet was found to have significant 
enhancements in resonant Raman scattering, as well as in 
spectral shifts of diffractively-coupled plasmon resonances. 
Plasmon resonances can be employed for investigating chemical properties of either graphene or another adjacent bulk surface, 
displaying the maturing capabilities of graphene-based plasmonics.\,\cite{poliNS,PNS01,PNS02,PNS03,yan} 
A junction between graphene and metallic contacts could also be used to 
design and fabricate a high-performance transistor. Consequently, exact knowledge
about the plasmon dispersions and their mode damping in hybrid devices based on the newly discovered 2D materials seems
absolutely necessary for the full development of these novel devices and the extension of their applications.\,\cite{rev1,rev2} 
\medskip

Historically, it is well known that the time evolution of a plasmon excitation
in a {\em closed} system (e.g., free-standing 2D layers) is determined by two-particle Green's functions in many-body theory, 
from which we are able to derive both the plasmon dispersion and the
plasmon dissipation (damping) rate.\,\cite{Gbook}
However, in an {\em open} system,\,\cite{dhh-2,dhh-3,dhh-4,dhh-5,dhh-6,dhh-7,dhh-8, Rec} the time evolution
for electronic excitations becomes much more involved since it depends strongly on the Coulomb interaction with the environment (e.g., electron reservoirs). 
As an example, the classical and quantum dynamical phenomena in open systems include tunnel-coupling to external electrodes,\,\cite{dhh-9} 
optical-cavity leakage to free space,\,\cite{dhh-10} and thermal coupling to heat baths.\,\cite{dhh-11} 
The coupling to an external reservoir is usually accompanied by extra dissipation channels. Many dynamical energy-dissipation theories for open systems
are based on the so-called Lindblad dissipative superoperator.\,\cite{dhh-12}
\medskip

In spite of the obvious advantages, such as being fast and tunable, in designing graphene-based devices, creating a sizable energy bandgap
has become an important issue for the practical use of graphene for transistors. The reason behind this issue is electrons in gapless graphene
may not be confined well by an electrostatic gate voltage or blocked by an energy barrier.\,\cite{Kl,ezawa} Scientists suggested
many approaches for opening an energy gap around $\backsimeq 0.1\,$eV by using various insulating substrates,\,\cite{gap2, gap3, gap4}
and graphene nanoribbons with quantized transverse wave vectors, or even by shining an intensive and polarized irradiation to dress electrons in graphene.\,\cite{kibis}
\medskip

In this respect, experimental implementation of a 2D lattice with a sizable spin-orbit
coupling seems to be an important advancement. Buckled structures, such as silicene and germanene in which
atoms are displaced out of the plane due to $sp^3$ hybridization, exhibit significant
asymmetry with respect to the $a-$ and $b-$sublattices. This leads to a new type of bandgap which is tunable by applying a perpendicular electric
field. Physically, silicene and other buckled honeycomb lattices have been modeled successfully by introducing the Kane-Mele type Hamiltonian\,\cite{KaneMele} 
with an intrinsic spin-orbit energy gap ($1.5 - 7.9\,$meV for silicene) to low-energy electrons. 
Systems meeting such requirements have already been realized experimentally.\,\cite{ezawa,ezawaprl,ezawa9prl} This includes 
recently synthesized germanene\,\cite{G1zhang,G2acun,G11li,G12davila,G13bampoulis,G14derivaz} with a considerably
larger spin-orbit coupling and the bandgap of $24-93\,$meV. The Hamiltonian, energy dispersion, and 
related electronic properties of germanene are qualitatively similar to those of silicene, although the Fermi velocity, 
the bandgap induced by spin-orbit coupling and the buckling height in germanene are still different in magnitudes. 
We, therefore, believe that our previous theory\,\cite{previous} for silicene can equally be applicable to Ge-based hybrid structures. 
\medskip

Germanene layers have been fabricated by molecular beam epitaxy on $\mathrm{Ag}(1\,1\,1)$ surfaces
through deposition on h-AlN and investigated 
by x-ray absorption spectroscopy.\,\cite{R2} 
Here, h-AlN was used to create an insulating buffer layer between germanene and its metal substrate. 
The measured lattice constant is in good agreement with the 
theoretical predictions for a free-standing Ge layer. In addition, a thorough experimental study for the density of states of germanene
which had been synthesized on Ge$/$Pt crystals at finite temperatures was performed by using a scanning-tunneling electron microscope.\,\cite{R3} 
The obtained virtually perfect linearly dependent density of states is clearly a proof of a 
2D Dirac system. Furthermore, Friedel oscillations were not observed, implying the possible Klein paradox within the considered system.
\medskip

Recently, there has been a number of reported studies on microscopic electronic properties, insulating regimes and 
topologically protected edge states within a certain range of applied electric fields, as well as spin- and valley-polarized quantum
Hall effects.\,\cite{Dru4,Dru5,Dru6,TabNicPRL,TabNicACDC,TabNicMagneto} Here, a crucial feature for a buckled structure is the occurrence 
of the topological-insulator (TI) properties when the external electrostatic field is relatively low and the resulting
field-related energy gap $\Delta_z$ becomes less than the intrinsic spin-orbit one. If the two gaps are equal, on the other hand,
the system turns into a spin-valley polarized metal (VSPM) since the gap for one of the two subbands will close. For a strong electric field, the system behaves just like a regular band insulator (BI).\,\cite{SilMain, ezawa}
One expects that the TI phase can show some unique electronic properties.\,\cite{Qi, Hasan} Indeed,
the properties of a TI, and the energy bandgap of Si as well, are experimentally found tunable\,\cite{R21} 
by an in-plane biaxial strain.
\medskip

In addition to silicene and germanene, another atomically thin 2D material with a great potential for applications in electronic devices
is $\mathrm{MoS}_{2}$ (monolayer molybdenum disulfide, or ML-MDS) which is a prototype of a metal dichalcogenide. The first-principle 
studies of the electron structure of this material has predicted a hybridization of the $d$-orbitals of molybdenum atoms with the $p$-orbitals of sulfur atoms,
giving rise to a two-band continuum model for  $\mathrm{MoS}_{2}$ monolayer.\,\cite{MoS01,MoS02,MoS03,Hab} This model further indicates the existence of two spin and valley degenerate subbands 
with a very large direct bandgap ($1.9\,$eV) and a strong spin-orbit interaction.\,\cite{xiao-prl,mos2}. It is 
important to note that the electronic properties of a ML-MDS are drastically different from its bulk samples with an indirect bandgap $\backsimeq 1.3\,$eV. 
Apart from the bandgap difference, the mobility of a single-layer $\mathrm{MoS}_{2}$ at 
the room-temperature exceeds $200\,$cm$^2$V$^{-1}$s$^{-1}$ and also acquires an ultralow standby power dissipation. Together with its direct bandgap, this makes 
ML-MDS an excellent candidate for next generation field-effect transistors. In strong contrast to the bulk, the $\mathrm{MoS}_{2}$ monolayer 
can emit light efficiently, and a possible high-temperature superfluidity, as a two-component Bose gas, has been predicted\,\cite{Oleg16}
in a system involving a transition metal dichalcogenide bilayer. Experimentally,
germanene has already been successfully synthesized on $\mathrm{MoS}_2$ substrate.\,\cite{R1} This leads to a huge advantage in comparison with the germanene synthesis on a metallic substrate
in which germanene is strongly hybridized with metals, leading to two parabolic bands, instead of linearly dispersing bands, at its six $K$ points.
\medskip

\begin{figure}
\centering
\includegraphics[width=0.49\textwidth]{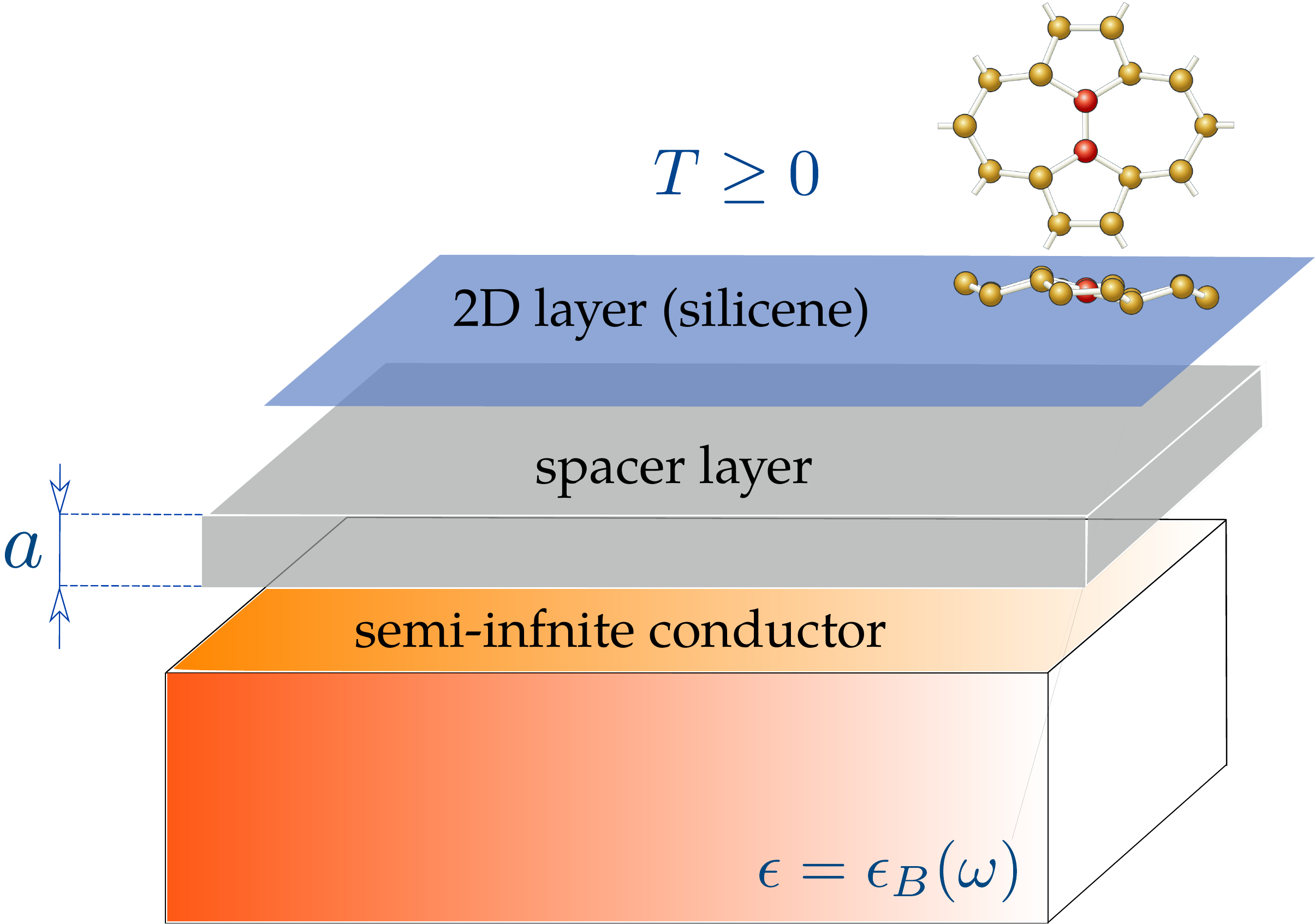}
\caption{Schematics of a hybrid plasmonic structure, including a 2D layer (silicene, 
germanene, MoS$_2$ etc.) interacting with a semi-infinite conducting substrate, separated from the layer by a distance $a$.}
\label{FIG:1}
\end{figure}

Our main focus in this paper is studying electron interactions in \textit{2D-material open systems} (2DMOS), i.e., a nanoscale hybrid structure of a 2D
layer (graphene with or without energy gap, silicene, germanene, a transition-metal dichalcogenide monolayer), as illustrated by Fig.\,\ref{FIG:1}. The most distinguished feature of 
such an open system is the dynamically-screened Coulomb interaction between electrons in graphene and the conducting substrate.\,\cite{ourPRB15} 
Screening is included by calculating the nonlocal frequency-dependent inverse dielectric function $\mc{K}({\bf r},{\bf r}';\,\omega)$ which is related to the regular
dielectric function $\epsilon({\bf r},\,{\bf r''};\,\omega)$ by $\int d^3{\bf r'}\, \mc{K}({\bf r},{\bf r'};\, \omega)\,\epsilon({\bf r'},{\bf r''};\,\omega)
=\delta({\bf r} - {\bf r''})$. The resonance occurring in $\mc{K}({\bf r},{\bf r}';\,\omega)$ corresponds to the nonlocal plasmon modes. 
The bare Coulomb potential $v({\bf r})$ in this open system would be screened, leading to 
$U_{\rm eff}({\bf r},\,\omega) = \int d^3{\bf r'}\,\mc{K}({\bf r},\, {\bf r'};\,\omega)\,v({\bf r'})$.
The mean-field formalism for nonlocal plasmon modes of a 2D layer interacting with a thick conductor
was discussed in Refs.\,[\onlinecite{Gbook,H0,Kouzakov,H1}]. This theory had given rise to a linear plasmon 
branch, which was later confirmed by an experiment.\,\cite{kram} Similar plasmon branches were also shown for 
double graphene layers at different temperatures.\,\cite{ourPRB15,coherent,myT,hbook,ourSREP} Furthermore, our previous work
studied the plasmon instability in the graphene double-layer system, predicting an instability-based terahertz emission,\,\cite{spiler} 
and nonlocal plasmons in a metal-graphene-metal encapsulated structure.\,\cite{gsandw}
\medskip

As it was reported before\,\cite{ourPRB15} the energy gap plays a crucial role on the nonlocal collective modes since it affects both
plasmon branches and Landau damping due to single-particle excitations.\,\cite{pavlo1} Taking
this into account, we focus on the effects of energy gaps and particle-hole modes (PHMs) in all three distinguishable insulating regimes of silicene, 
i.e., TI, VSPM and regular BI.\,\cite{ezawa,ezawaprl,ezawa9prl,SilMain}
Here, both the energy gaps and PHMs could be independently tuned by applying an electric field perpendicular to the silicene layer.
\medskip

The rest of the paper is organized as follows. In Sec.\ \ref{sect2}, we first study nonlocal plasmon excitations in the silicene system.
Our numerical results show plasmon modes and their damping as functions of frequencies and wave numbers for different parameters including Fermi
energies, spin-orbit and sublattice asymmetry bandgaps, and various surface-plasmon frequencies and coupling strengths. In some limiting cases, analytical results for the plasmon modes are presented
and analyzed for their dependencies on sample structure parameters. In Sec.\ \ref{sect3}, we further explore the
band-energy dispersions and the electronic states of molybdenum disulfide, indicating relevance to recently discovered group \rom{4} dichalcogenides. 
For these materials, crucial analytical results are obtained for the wave functions, overlap factor and Fermi energy,
which have not yet been addressed adequately in the literature. Additionally, we also investigate nonlocal plasmon branches
in the long-wavelength limit and demonstrate how they are affected by the mismatch of $n$ and $p$ doping types and densities. Finally, concluding remarks and discussion
of our numerical results in this paper are presented in Sec.\ \ref{sect4}.

\section{Hybrid plasmon modes and damping in open silicene systems}
\label{sect2}

In this section, we discuss hybrid-plasmon dynamics in a silicene open system at low temperatures, and we will address the molybdenum-disulfide open system in the next section.
For a single silicene layer, the low-energy Hamiltonian\,\cite{SilMain,EzawaR,TN30,ezawa,ezawaprl} at the corners of the first Brillouin zone is

\begin{equation}
\hat{\mbb{H}}_{4\times 4} = \hbar v_F \left( 
\xi k_x \hat{\tau}_x + k_y \hat{\tau}_y
\right) \otimes \hat{\mbb{I}}_{2\times 2} - \xi \Delta_{SO} \hat{\sigma}_z \otimes \hat{\tau}_z +
\Delta_z \hat{\tau}_z \otimes \hat{\mbb{I}}_{2\times 2}\,\ ,
\label{matrix1}
\end{equation}
where $\Delta_{SO}$ is the intrinsic spin-orbit energy gap, $\Delta_z\propto\mc{E}_\bot$ represents the field-dependent sublattice asymmetry bandgap with $\mc{E}_\bot$ as an applied electric field perpendicular 
to the lattice, $\hat{\mbb{I}}_{2\times 2}$ is the unit matrix, $\hat{\tau}_{x,y,z}$ and $\hat{\sigma}_{x,y,z}$ are the $2\times 2$ Pauli matrices determining, respectively, the
electron spin and valley pseudospin states of the system, $\xi = \pm 1$ labels two inequivalent $K$ and $K'$ valleys, and the silicene Fermi velocity $v_F$ is just half of that in graphene.
\medskip

This Hamiltonian in Eq.\,(\ref{matrix1}) can be cast into block-diagonal form with two $2\times 2$ matrices given by

\begin{equation}
\hat{\mbb{H}}_{\xi,\sigma} = \left[\begin{matrix}
- \xi \sigma \Delta_{SO} + \Delta_z & \hbar v_F (\xi k_x - i k_y) \\
\hbar v_F (\xi k_x + i k_y) & \xi \sigma \Delta_{SO} - \Delta_z 
\end{matrix}\right]\, ,
\label{mainHamS}
\end{equation}
where $\sigma=\pm 1$ is the spin eigenvalue of $\hat{\sigma}_z$ and $\xi=\pm 1$ is the valley indices. The energy dispersions, $\mbb{E}_{\xi,\sigma}(k)$, associated with Eq.\,(\ref{mainHamS}) are 

\begin{equation}
\pm\mbb{E}_{\xi,\sigma}(k) = \pm \sqrt{
\hbar^2 v_F^2k^2+\Delta_{\xi,\sigma}^2}\, ,
\label{esil}
\end{equation}
which represent a pair of spin-dependent energy subbands for each valley and have \textit{two} corresponding
non-equivalent bandgaps $\Delta_{\xi,\sigma} \equiv \vert \Delta_{SO} - \xi \sigma \Delta_z \vert = \vert \Delta_{SO} \pm \Delta_z \vert $. 
The positive (negative) sign in Eq.\,(\ref{esil}) is for electron (hole) states. For simplicity, we, therefore,
introduce the notations, $\Delta_{>}  = \Delta_{SO}+\Delta_z $ and 
$\Delta_{<} = \vert \Delta_{SO} - \Delta_z \vert $, for these two unequivalent bandgaps.
The key issues in this section are obtaining spectra of hybrid-plasmon excitations in 2DMOS with new ingredients $\Delta_{>,<}$
and calculating screened Coulomb couplings of electrons in silicene layers to 
an adjacent semi-infinite bulk plasma.
\medskip

It is known that the coupling of electrons in 2D materials to other conduction electrons in 2DMOS will change the plasmon-mode dispersion. Since the damping region is still decided by the 
PHMs in 2D materials, this will lead to a modification to the damping of plasmon modes in 2DMOS.
According to Refs.\,[\onlinecite{Gbook,H0,Kouzakov,H1}], the Fourier-transformed nonlocal composite inverse dielectric function can be determined by 

\begin{equation}
\mathcal{K}(z_1,z_2;\,q,\omega)= K_S(z_1,z_2;\,q,\omega)+\Pi_0(q,\,\omega)\,\frac{K_S(a,z_2;\,q,\omega)}{\mathbb{S}_{C}(q,\,\omega)}
\left\{ \int_{-\infty}^\infty dz^\prime\,K_S(z_1,z^\prime;\,q,\omega)\,v_c(q,\,|z^\prime-a|) \right\}\, .
\label{eq:GG-inv}
\end{equation}
In Eq.\,(\ref{eq:GG-inv}), $\Pi_0(q,\,\omega)$ is the electron polarizability of silicene (explicitly given below),
the interaction of silicene with the substrate is included in the second term,
$a$ is the separation of the silicene layer from the conducting surface, and $v_c(q,\,|z-z^\prime|)=(e^2/2\epsilon_0\epsilon_r)\,\exp(-q|z-z^\prime|)$
with $\epsilon_r$ as the average dielectric constant of silicene and spacer layer. Moreover,
$K_S(z_1,z_2;\,q,\omega)$ represents the local inverse dielectric function of the conducting substrate, expressed as

\begin{eqnarray}
K_S(z,z^\prime;\,q,\omega) &&  = \, \theta(z)\left\{ \delta(|z|-z^\prime)+\delta(z^\prime)\,
e^{-q|z|} \left[ \frac{1-\epsilon_B(\omega)}{1+\epsilon_B(\omega)} \right]  \right\}
\nonumber\\
&& + \, \theta(-z) \left\{ \frac{\delta(|z|+z^\prime)}{\epsilon_B(\omega)}+\delta(z^\prime)\,
e^{-q|z|}\, \frac{1}{\epsilon_B(\omega)} \left[ \frac{\epsilon_B
(\omega)-1}{\epsilon_B(\omega)+1} \right]  \right\} \ ,
\label{pol}
\end{eqnarray}
where $\theta(z)$ is a unit-step function, $z>0$ ($z<0$) corresponds to the spacer layer (conductor),
separated by the surface at $z=0$. We have also assumed a Drude model in Eq.\,(\ref{pol}) for the substrate dielectric function
$\epsilon_B(\omega)= 1 - \Omega_p^2/\omega^2$, where $\Omega_p = \sqrt{n_0e^2/\epsilon_0 \epsilon_b m^*}$
is the bulk-plasma frequency. Here, $\Omega_p$ depends on the electron concentration $n_0$, substrate dielectric constant $\epsilon_b$, 
the effective mass $m^*$ of electrons, and it can vary in a very large range from ultra-violet (metals) down to infrared or even terahertz (doped semiconductors) frequencies.
The use of the Drude model in Eq.\,(\ref{pol}) can be justified by a short screening length for high electron concentrations in bulk materials.
\medskip

Finally, we are in a position to calculate the hybrid-plasmon modes in 2DMOS.
The plasmon dispersions for a single silicene layer can be obtained from the dielectric-function equation: $\varepsilon(q,\,\omega) = 1 - (2\pi\alpha/q)\,\Pi_0(q,\,\omega) = 0$,
where $\alpha=e^2/4\pi\epsilon_0\epsilon_r$. For 2DMOS, on the other hand, $\varepsilon(q,\,\omega)$
should be replaced by the so-called ``dispersion factor'' $\mathbb{S}_C(q,\,\omega)$, which appears in Eq.\,(\ref{eq:GG-inv}) and is calculated as

\begin{equation}
\mathbb{S}_C(q,\,\omega) = 1 - \left(\frac{2\pi\alpha}{q}\right)\Pi_0(q,\,\omega) \left[  
1 + \texttt{e}^{-2qa} \,\frac{\Omega_p^2}{2\omega^2-\Omega_p^2}
\right]\, .
\label{sc}
\end{equation}
This verifies that the plasmon dispersions in 2DMOS will indeed be modified by coupling to other conduction electrons ($\Omega_p\neq 0$). 
\medskip

Since the Coulomb coupling between electrons in different ($K$ and $K'$) valleys involves two uncompensated
very large lattice wave numbers, the resulting electron interaction becomes negligible in comparison with those of electrons within the same valley.
Consequently, by ignoring inter-valley Coulomb coupling and using the one-loop approximation\,\cite{SilMain} for silicene, we find

\begin{equation}
\Pi_0(q,\,\omega) = \sum\limits_{\beta =>,<}\,\Pi_0(q,\omega;\,\Delta_{\beta}) \, ,
\label{sum}
\end{equation}
and for each subband we get

\begin{equation}
\Pi_0(q,\omega;\,\Delta_{\beta}) = \frac{1}{4 \pi^2} \int d^2{\bf k} \sum_{s,s'=\pm 1}\,\left[ 1 + ss'\,
\frac{(\hbar v_F)^2{\bf k \cdot ({\bf k} + {\bf q})} + \Delta_\beta^2 }{ \mbb{E}_{\beta}(k)\,\mbb{E}_{\beta}(|{ \bf k} + {\bf q}|)} \right]
\frac{f_0[s\mbb{E}_{\beta}(k)] - f_0[s'\mbb{E}_{\beta}(|{ \bf k} + {\bf q}|)]}
{s\mbb{E}_{\beta}(k)-s'\mbb{E}_{\beta}(|{ \bf k} + {\bf q}|)-\hbar\omega-i0^+}\, ,
\label{pi1p}
\end{equation}
where $s,\,s' = \pm 1$ denote electron and hole states, respectively, $f_0(\mbb{E})=\theta(\mbb{E}-E_F)$ at zero temperature and $E_F$ is the Fermi energy of electrons in silicene.
\medskip

\subsection{Approximate analytical results}

In the long-wavelength limit $q\ll k^\beta_F$ ($k^\beta_F$ is the Fermi wave number for each subband), the bare bubble polarization function for $E_F>\Delta_>$ is obtained as\,\cite{sensarma}
 
\begin{equation}
\Pi_0(q,\,\omega) = \frac{1}{\pi}\sum_{\beta=>,<}\,k^\beta_F\,\left|\frac{\partial\mbb{E}_\beta(k)}{\partial k}\right|_{k = k^\beta_F} \, \,
\frac{q^2}{\hbar^2\omega^2}
=\frac{E_F}{\pi} \left(2 - \frac{\Delta_<^2}{E_F^2} - \frac{\Delta_>^2}{E_F^2} \right) \, \frac{q^2}{\hbar^2\omega^2}\, ,
\label{longw}
\end{equation}
where $E_F=\sqrt{(\hbar v_Fk^\beta_F)^2+\Delta^2_\beta}$, $k_F^\beta=\sqrt{2\pi \rho_\beta}$, and $\rho_\beta$ is the electron areal density for each subband.
Therefore, from $\varepsilon(q,\,\omega)=0$ for a single silicene layer, we get the following plasmon branch

\begin{equation}
\omega_p^2(q)  = \frac{4\alpha}{\hbar^2E_F}\left( E_F^2 - \frac{\Delta_>^2 + \Delta_<^2}{2} \right) q \equiv \mbb{G} \, q\, ,
\label{mode}
\end{equation}
where, for convenience, we introduce a coefficient $\mbb{G}\equiv \mbb{G}(E_F,\,\Delta_{\beta})$.
\medskip

It is important to note that the Fermi energy $E_F$ for silicene is fixed by the total electron areal density $\rho_0$ through

\begin{equation}
E_F^2 - \frac{1}{2} \left( \Delta_>^2 + \Delta_<^2 \right)=(\hbar v_F)^2\pi(\rho_>+\rho_<)\equiv(\hbar v_F)^2\pi\rho_0  \, ,
\end{equation}
and if $\rho_0$ is small, we can further approximately obtain

\begin{equation}
E_F \backsimeq \sqrt{\frac{\Delta_>^2 + \Delta_<^2}{2}} + \frac{(\hbar v_F)^2\pi\rho_0}{\sqrt{2 \left( \Delta_>^2 + \Delta_<^2 \right)}} \, .
\label{restraint}
\end{equation}
This leads to $\mbb{G}\backsimeq 4\sqrt{2}\alpha v_F^2\pi\rho_0/(\sqrt{\Delta_>^2 + \Delta_<^2})$.
For gapped graphene, we have $\Delta_< = \Delta_> = \Delta$, and Eq.\,(\ref{restraint}) gives rise to $ E_F - \Delta \backsimeq
(\hbar v_F)^2\,\pi\rho_0/(2\Delta)$. In this case, we get $\mbb{G}\backsimeq 4\alpha v_F^2\pi\rho_0/\Delta$ and from Eq.\,(\ref{mode}) we find $\omega_p(q)\sim\sqrt{\rho_0q}$.
Actually, such a scaling relation holds true for all 2D materials except for gapless graphene which yields $\omega_p(q)\sim\sqrt{\rho^{1/2}_0q}$.
Additionally, in contrast to Eq.\,(\ref{mode}), if $E_F < \Delta_>$ with an unoccupied upper subband, we obtain the plasmon mode

\begin{equation}
\omega_p^2(q)= \frac{2\alpha E_F}{\hbar^2}\left( 1 - \frac{\Delta_<^2}{E_F^2}\right) q\, . 
\end{equation}
\medskip

In the above discussion, we are only restricted to the plasmon mode for a stand-alone silicene. For the 2DMOS, on the other hand, when both subbands are occupied, the plasmon modes are determined by Eqs.\,(\ref{sc}) and (\ref{longw}), yielding

\begin{equation}
1 - \frac{\mbb{G}q}{\omega^2}\left(1 + \texttt{e}^{-2qa}\,\frac{\Omega_p^2}{2\omega^2 - \Omega_p^2}\right)  = 0 \, ,
\end{equation}
which gives rise to the following bi-quadratic equation

\begin{equation}
2 \left(\frac{\omega^2}{\Omega^2_p} \right)^2 - \left(1 + \frac{2\mbb{G}q}{\Omega_p^2}\right) \left(\frac{\omega^2}{\Omega^2_p} \right) + 
\frac{\mbb{G} q}{\Omega_p^2} \left( 1 - \tet{e}^{-2qa} \right) = 0 \, .
\end{equation}
Its two solutions are simply given by

\begin{equation}
\frac{4\omega^2_{p,\pm}}{\Omega^2_p}=\left(1 + \frac{2\mbb{G}q}{\Omega_p^2}\right)\pm\left[\left(1 + \frac{2\mbb{G}q}{\Omega_p^2}\right)^2-\frac{8\mbb{G} q}{\Omega_p^2} \left( 1 - \tet{e}^{-2qa} \right)\right]^{1/2}\, ,
\end{equation}
where the sign $+$ ($-$) corresponds to the in-phase (out-of-phase) plasmon mode.
For the strong-coupling regime with $qa\ll 1$ and in the long-wavelength limit, the split hybrid plasmon modes are found to be

\begin{eqnarray}
\nonumber
&& \omega_{p,+}(q) \backsimeq \frac{\Omega_p}{\sqrt{2}} + \frac{\mbb{G}q}{\sqrt{2}\Omega_p}- \frac{\mbb{G}+4a\Omega_p^2}{2\sqrt{2}\,\Omega_p^3}\,\mbb{G}q^2 +{\cal O}(q^3) \, ,
\\
&& \omega_{p,-}(q) \backsimeq q\,\sqrt{2a\mbb{G}}-\frac{ \sqrt{2a\mbb{G}}}{\Omega_p^2}\, \mbb{G}q^2+{\cal O}(q^3)\, .
\label{sl}
\end{eqnarray}
\medskip

In Eq.\,(\ref{sl}), the linear dispersions and their prefactor scalings are the same as those for graphene.\,\cite{ourPRB15} 
However, the two independent bandgaps, $\Delta_{SO}$ and $\Delta_z$, play unique roles in shaping the hybrid-plasmon branches when the damping from different PHMs is considered. 
It is found that the outer PHM's boundaries are only determined by $\Delta_{<}$ and the two hybrid-plasmon group velocities (slopes) are proportional to $\mbb{G}$ and $\sqrt{\mbb{G}}$.
These two group velocities drop to zero as $\Delta_{SO}$ and $\Delta_z$ increase since $\mbb{G}\backsimeq 4\sqrt{2}\alpha v_F^2\pi\rho_0/\sqrt{\Delta_>^2 + \Delta_<^2}$ for low doping. 
If a proper applied electric field is chosen, the VSPM phase can be reached with $\Delta_{<}=0$. For $\Delta_{<}>0$, on the other hand, 
we know the plasmon frequencies for both gapped graphene\,\cite{pavlo1} and silicene\,\cite{SilMain} are reduced by finite $\Delta_{<}$. Meanwhile,
these plasmon branches will enter into a gap region between the interband and intraband PHMs. As a result, we find that 
both the damping-free plasmon regions and the plasmon group velocities can be controlled independently by $\Delta_{SO}$ and $\Delta_z$. This requirement can be fulfilled by scanning an external electric field, 
even for a fixed spin-orbit interaction strength, leading to distinctive behaviors for TI and BI phases. 
\medskip

Moreover, the plasmon group velocity associated with the $\omega_{p,-}(q)$ mode in 2DMOS depends on $\sqrt{a}$ in the strong-coupling regime. 
However, in the weak-coupling regime with $qa\gg 1$, this plasmon mode becomes proportional to $\sqrt{q}$, as shown in Eq.\,(\ref{mode}) for a single silicene layer. 
Meanwhile, the plasmon group velocity for the $\omega_{p,+}(q)$ mode, which is independent of $a$, approaches zero in the weak-coupling regime.
 
\begin{figure}
\centering
\includegraphics[width=0.49\textwidth]{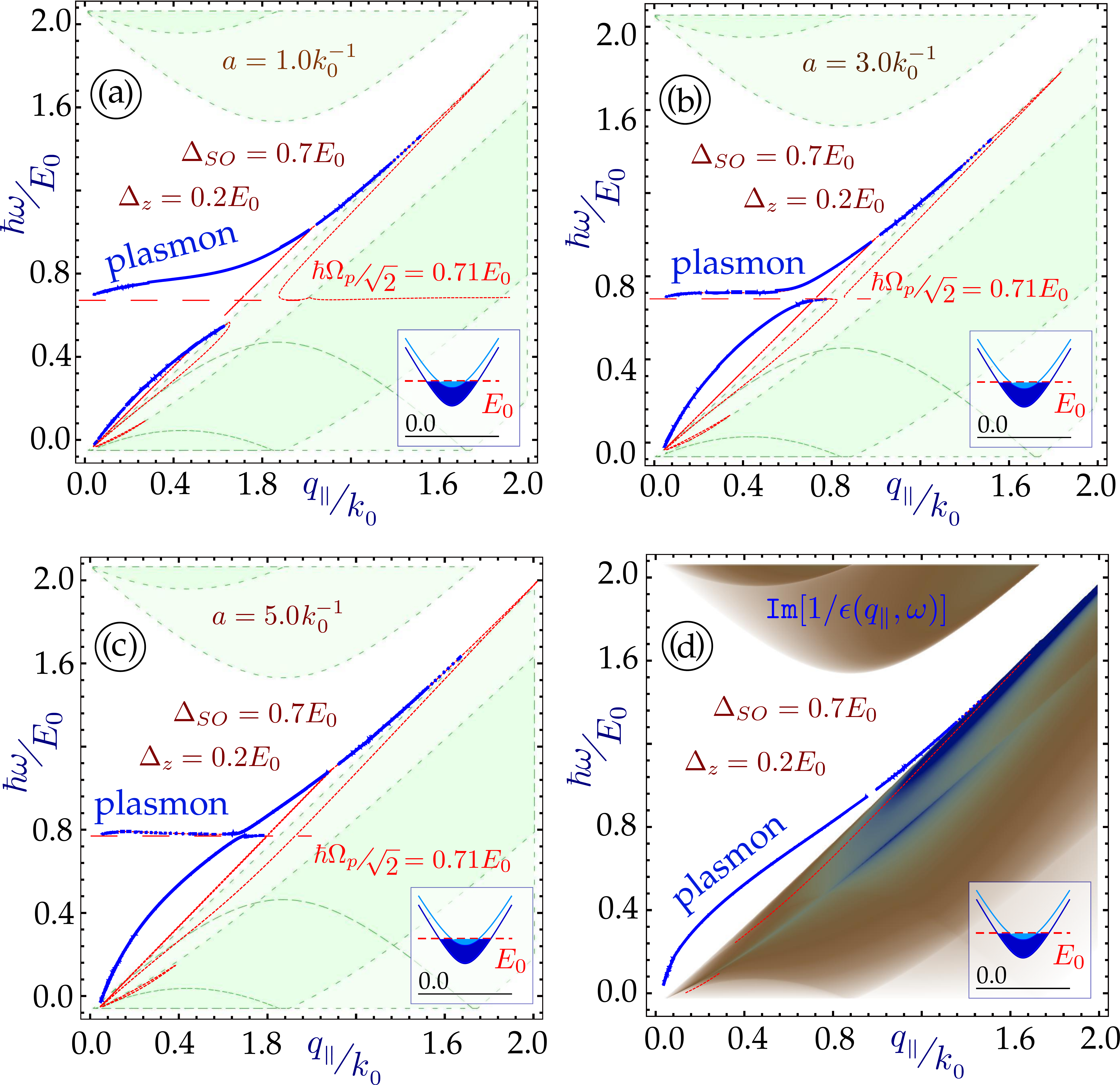}
\caption{Numerical results for the hybrid-plasmon branches in 2DMOS with $\Delta_{SO}/E_0=0.7$ and $\Delta_{z}/E_0=0.2$ ($\Delta_</E_0=0.5$).
Plots $(a)$-$(c)$ correspond to various cases with $k_0a=1.0$, $3.0$ and $5.0$. Here, the plasma energy $\hbar\Omega_p/E_0=1.0$.
In all panels, the blue solid curves are obtained from $|\mathbb{S}_c(q,\,\omega)|=0$, while the red short-dashed curves
are from $\text{Re}\left[\mathbb{S}_c(q,\,\omega)\right]=0$, 
demonstrating both undamped and damped plasmon branches. The PHM regions are depicted by partially transparent green areas enclosed by dashed-curve boundaries.  
Panel $(d)$ gives the density plot for the energy loss function $\text{Im}\left[1/\epsilon(q,\,\omega)\right]$ of free-standing silicene. 
The populations of two subbands are shown in the inset.}
\label{FIG:2}
\end{figure}

\begin{figure}
\centering
\includegraphics[width=0.49\textwidth]{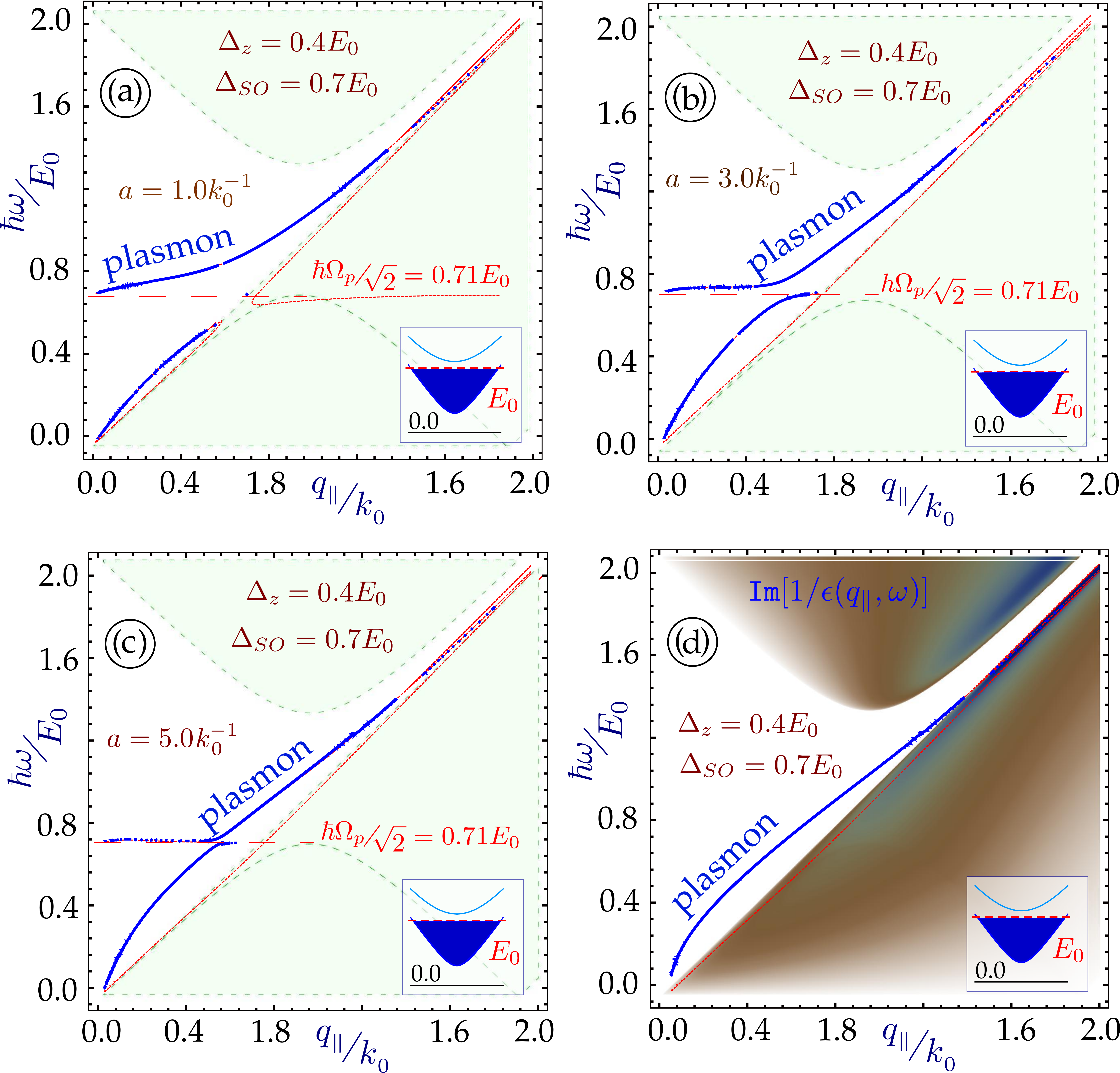}
\caption{Numerical results for the hybrid-plasmon branches in 2DMOS with $\Delta_{SO}/E_0=0.7$ and $\Delta_{z}/E_0=0.4$ ($\Delta_</E_0=0.3$).
Plots $(a)$-$(c)$ correspond to $k_0a=1.0$, $3.0$ and $5.0$, respectively. Here, $\hbar\Omega_p/E_0=1.0$.
In all panels, the blue solid curves are for the undamped plasmon modes, while the red short-dashed curves are for the damped plasmon modes. 
The PHM regions are depicted by partially transparent green areas enclosed by dashed-curve boundaries.  
Panel $(d)$ gives the density plot for $\text{Im}\left[1/\epsilon(q,\,\omega)\right]$ of a single silicene layer. 
The populations of two subbands are shown in the inset.}
\label{FIG:3}
\end{figure}

\begin{figure}
\centering
\includegraphics[width=0.49\textwidth]{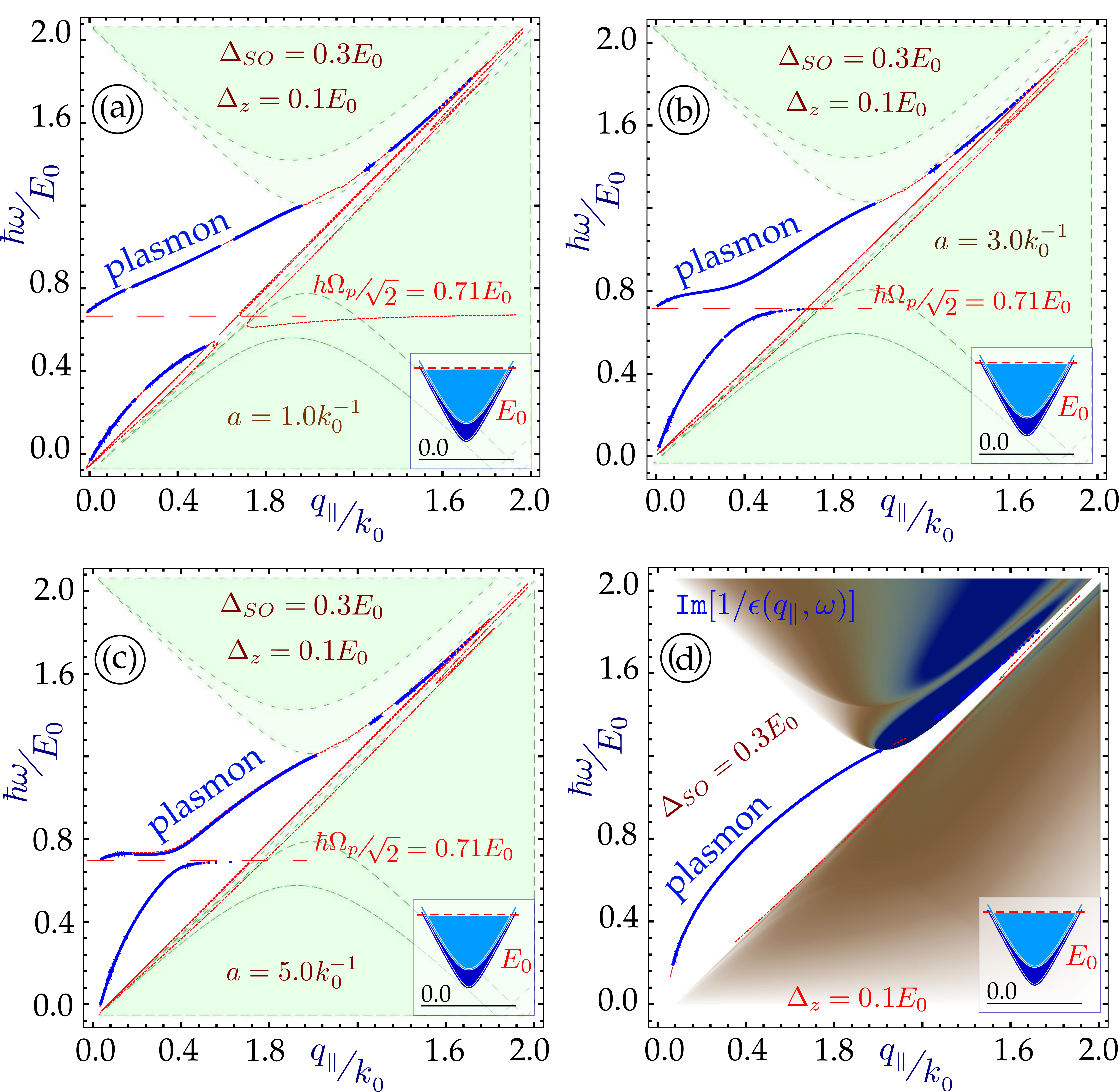}
\caption{Numerical results for the hybrid-plasmon branches in 2DMOS with $\Delta_{SO}/E_0=0.3$ and $\Delta_{z}/E_0=0.1$ ($\Delta_</E_0=0.2$).
Plots $(a)$-$(c)$ correspond to $k_Fa=1.0$, $3.0$ and $5.0$, respectively. Here, $\hbar\Omega_p/E_0=1.0$.
In all panels, the blue solid curves are for the undamped plasmon modes, while the red short-dashed curves for the undamped plasmon modes. 
The PHM regions are depicted by partially transparent green areas enclosed by dashed-curve boundaries.  
Panel $(d)$ gives the density plot for $\text{Im}\left[1/\epsilon(q,\,\omega)\right]$ of a single silicene layer. 
The populations of two subbands are shown in the inset.}
\label{FIG:4}
\end{figure}

\begin{figure}
\centering
\includegraphics[width=0.49\textwidth]{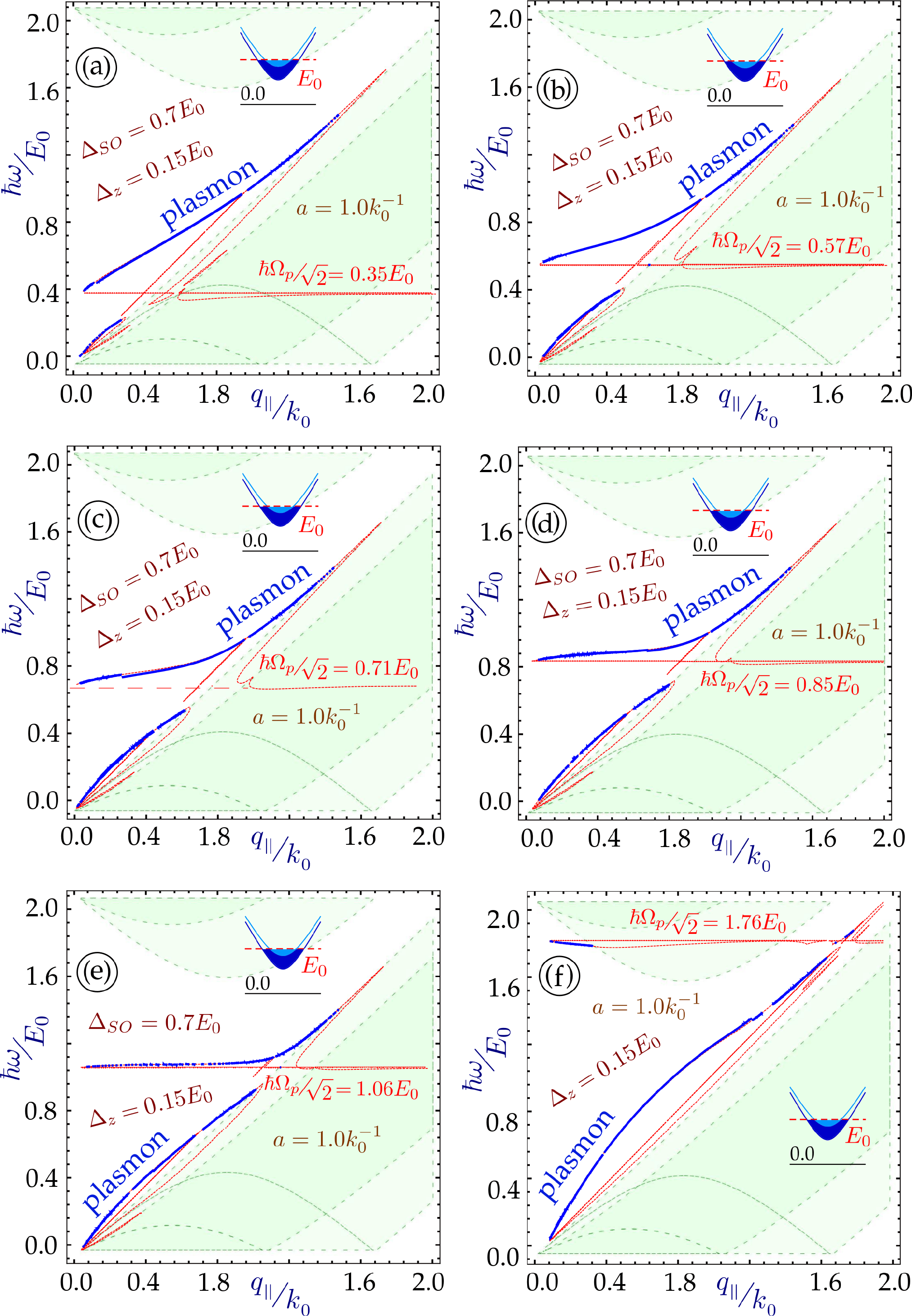}
\caption{Numerical results for the hybrid-plasmon branches in 2DMOS with $\Delta_{SO}/E_0=0.7$ and $\Delta_{z}/E_0=0.15$ ($\Delta_</E_0=0.55$).
Plots $(a)$-$(f)$ correspond to $\hbar\Omega_p/E_0=0.5$, $0.8$, $1.0$, $1.2$, $1.5$ and $2.5$, respectively. Here, $k_0a=1.0$.
In all panels, the blue solid curves are for the undamped plasmon modes, while the red short-dashed curves for the undamped plasmon modes. 
The PHM regions are depicted by partially transparent green areas enclosed by with dashed-curve boundaries.  
The populations of two subbands are shown in the inset.}
\label{FIG:5}
\end{figure}

\begin{figure}
\centering
\includegraphics[width=0.49\textwidth]{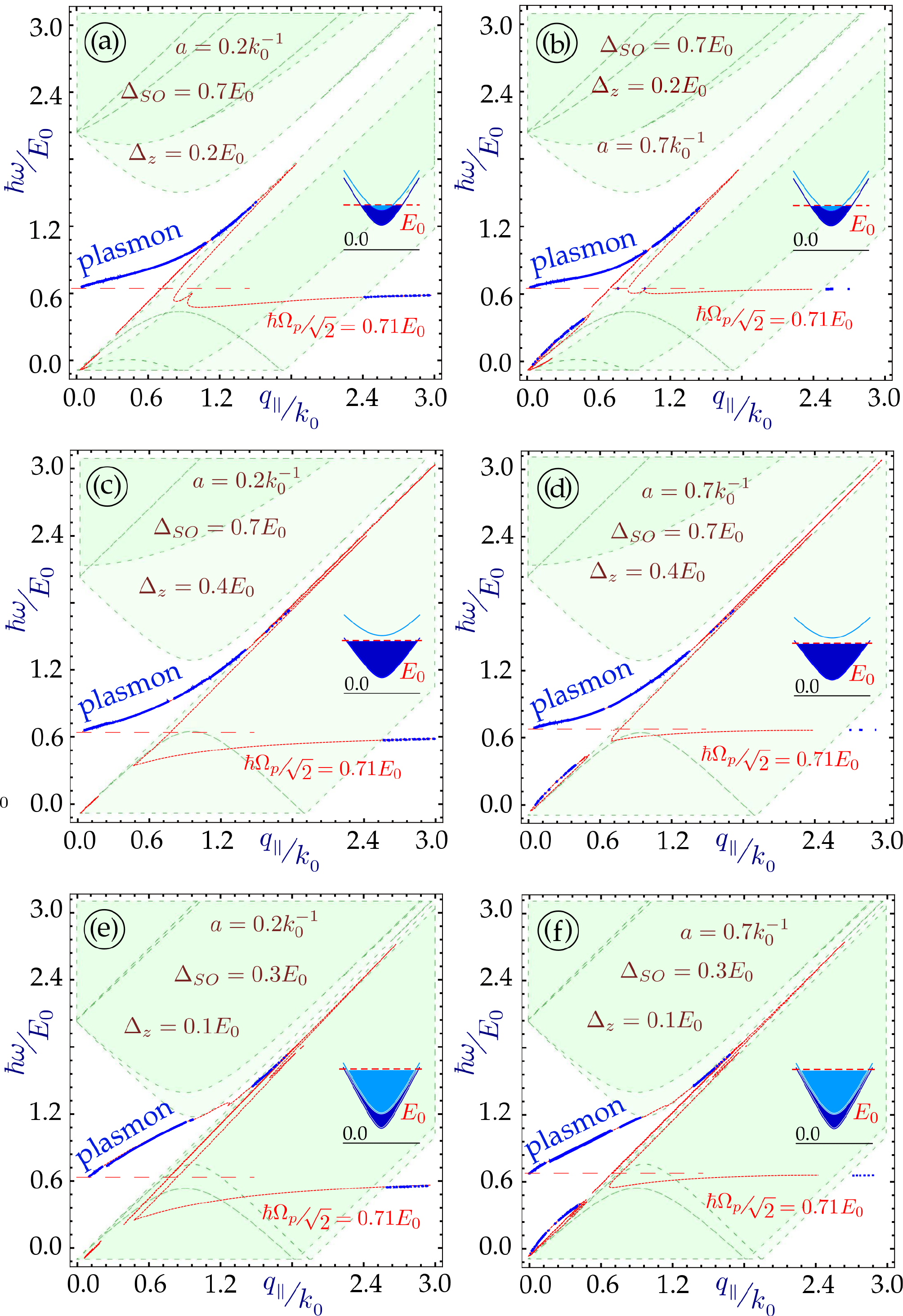}
\caption{Numerical results for the hybrid-plasmon branches in 2DMOS. Here, plots $(a)$, $(c)$ and $(e)$ are for 
$k_0a = 0.2$, while plots $(b)$, $(d)$ and $(f)$ for $k_0a = 0.7$.
In addition, we assume $\Delta_{SO}/E_0=0.7$ and $\Delta_{z}/E_0=0.2$ ($\Delta_</E_0=0.5$) in $(a)$ and $(b)$, 
$\Delta_{SO}/E_0=0.7$ and $\Delta_{z}/0_F=0.4$ ($\Delta_</E_0=0.3$) in $(c)$ and $(d)$,
and $\Delta_{SO}/E_0=0.3$ and $\Delta_{z}/E_0=0.1$ ($\Delta_</E_0=0.2$) in $(e)$ and $(f)$.
In all panels, the blue solid curves are for the undamped plasmon modes, while the red short-dashed curves for the undamped plasmon modes. 
The PHM regions are depicted by partially transparent green areas enclosed by the dashed-curve boundaries.  
The populations of two subbands are shown in different insets.}
\label{FIG:6}
\end{figure}

\subsection{Full numerical solutions}

For our numerical results presented in Figs.\,\ref{FIG:2}-\ref{FIG:6}, we use the scale $E_0$ for the energy and the scale $k_0$ for the wave number $q$, where $E_0=\hbar v_Fk_0$,
$k_0=\sqrt{\pi\rho_0}$ with $\rho_0$ as the total doped electron areal density. Here, the constant value $\rho_0=10^{15}\,$cm$^{-2}$ is given for these five figures.
\medskip

The features of the hybrid plasmon modes beyond the long-wavelength limit could be explored numerically 
for all possible values of the energy bandgaps based on the exact calculation of the polarization function for a silicene layer\,\cite{SilMain} 
and the use of Eq.\,(\ref{sc}). Here, two subbands can be selectively populated by controlling $E_F$ or $\rho_0$. Furthermore, the coupling of electrons to the surface of a semi-infinite conductor in 2DMOS can 
also be tuned by choosing the separation of the silicene layer from the bulk surface. 
\medskip

We first consider a case with a relatively large minimal bandgap $\Delta_</E_F = 0.5$ and both subbands occupied. The numerical results for this case  
are presented in Fig.\,\ref{FIG:2}, where both the dispersion and undamped extension of the lower acoustic-like branch mainly
depend on the separation $a$. The anticrossing of two hybrid plasmon modes can be seen most clearly in Fig.\,\ref{FIG:2}($b$) with $k_Fa=3.0$.
In comparison with the plasmon damping for free-standing slicene in Fig.\,\ref{FIG:2}($d$), 
the upper optical-like branch is free from damping into the main diagonal ($\omega = v_F q$, intraband PHM) until exceeding a relatively large critical wave number, as shown in Figs.\,\ref{FIG:2}($a$)-\ref{FIG:2}($c$).
For $\Delta_</E_F = 0.3$ in Fig.\,\ref{FIG:3}, where only the lower subband is occupied, the plasmon-mode dispersions
are found to be similar to gapped graphene in Ref.\,[\onlinecite{ourPRB15}]. For $k_Fa=5.0$ in Fig.\,\ref{FIG:3}($c$), 
the anticrossing feature becomes almost indistinguishable, and the lower branch approaches that of free-standing slicene in Fig.\,\ref{FIG:3}($d$). 
\medskip

The situation with an even smaller bandgap $\Delta_</E_F = 0.2$ and two occupied subbands is presented in Fig.\,\ref{FIG:4}. Here, we find an unusual feature 
that the upper branch damps into both intraband and interband PHM regions at different wave numbers. 
For a free-standing silicene sample in Fig.\,\ref{FIG:4}($d$), however, the damping always occurs at one of the PHM boundaries, and the increase of bandgap 
makes it more favorable for the plasmon-mode damping to occur at the interband PHM region. In addition, we also find that the anticrossing feature 
becomes more significant in the weak-coupling regime, as displayed in Fig.\,\ref{FIG:4}($c$).
\medskip

It is known that the external parameter in the 2DMOS, i.e., the plasma frequency $\Omega_p$, can greatly affect the damping of the hybrid plasmon modes. 
Our results for different values of $\Omega_p$ are shown in Fig.\,\ref{FIG:5}. It is very surprising to see from Figs.\,\ref{FIG:5}($a$)-\ref{FIG:5}($e$) that the damping-free range of the lower branch will depend on
$\Omega_p$ but not on the other internal parameters, such as $a$, $\Delta_<$ and $\rho_0$. From Fig.\,\ref{FIG:5}($f$), on the other hand, we observe that 
the upper branch could be doubly damped by both intraband and interband PHM regions, which also exists for a gapped graphene open system. 
\medskip

It is reasonable to expect that the open-system damping effect will become more significant if the conductor surface stays closer to the silicene layer.
The numerical results for the plasmon dispersions with much smaller separations $a$ are presented in Fig.\,\ref{FIG:6}, from which we reproduce a recent experimentally confirmed 
effect in graphene,\,\cite{poli1,poli2,poli3,hbook}  i.e., the acoustic-like plasmon branch will be highly damped in the long-wavelength limit as $a<0.5\,$nm. 
This damping effect can be found for all considered cases in Figs.\,\ref{FIG:6}($a$)-\ref{FIG:6}($f$) with various energy gaps. 
It is interesting to note from Fig.\,\ref{FIG:6} that the group velocity of the upper branch almost does not depend on the separation
$a$, but strongly depends on the energy gap $\Delta_<$. 

\section{Hybrid plasmon modes and damping in open molybdenum-disulfide systems}
\label{sect3}

The two-band model Hamiltonian for molybdenum disulfide, as well as for most other transition-metal dichalcogenides,
next to the two inequivalent $K$ and $K'$ valley points can be written as\,\cite{xiao-prl,mos2}

\begin{equation}
\hat{\mbb{H}}_{d}^{\tau,s} = \left( 
\frac{1}{2}\,\tau s \, \lambda_0 + \frac{\hbar^2 k^2}{4 m_e} \alpha 
\right) \hat{\mbb{I}}_{2\times 2}
+
\left(
\frac{\Delta}{2} - \frac{1}{2}\,\tau s \, \lambda_0 + \frac{\hbar^2 k^2}{4 m_e} \beta 
\right) \hat{\sigma}_z + 
t_0a_0\,\hat{\boldsymbol{\mathrm{\Sigma}}}_\tau \cdot {\bf k}
\label{mosham}
\end{equation}
where $\tau=\pm 1$ and $s=\pm 1$ are the valley and spin indices, $\Delta = 1.9\,$eV is the main energy bandgap, $\lambda_0=0.042\,\Delta$ 
is the spin-orbit coupling parameter, $m_e$ represents the free electron mass, $\hat{\boldsymbol{\mathrm{\Sigma}}}_\tau$ are the Pauli matrices for valley pseudospins,
$t_0 = 0.884\,\Delta$ is the electron hopping parameter, and $a_0=1.843\,$\AA\ which is obtained from the Mo$-$S 
atom-atom bond length $2.43\,$\AA. Even though $\lambda_0 \ll \Delta$, the spin-orbit interaction is not negligible, which is reflected 
in the spin-resolved energy subbands and in the absence of spin degeneracy. 
In Eq.\,(\ref{mosham}), we use $\alpha =2.21 = 5.140\,\beta$ and we find that $t_0a_0= 4.95\times 10^{-29}\,$J$\cdot$m plays a role of the Fermi velocity
and is equal to $0.472$ of the $\hbar v_F$ factor for graphene.  
Moreover, we neglect the trigonal warping term $t_1 a_0^2\,(\hat{\boldsymbol{\mathrm{\Sigma}}}_\tau\cdot {\bf k})\,\hat{\sigma}_x(\hat{\boldsymbol{\mathrm{\Sigma}}}_\tau\cdot {\bf k})$, which leads to the 
slight anisotropy of the energy for our whole study since $t_1 = 0.1\,$eV$= 0.053\,\Delta$ does not represent a considerable effect
on the electronic states. Consequently, the considered hybrid plasmon are also isotropic. 
\medskip

It is easy to verify that the Hamiltonian in Eq.\,(\ref{mosham}) is equivalent to that of gapped graphene with a $k-$dependent ``gap'' term,
$\Delta_0^{\tau,s}(k) = \Delta / 2 - \tau s \, \lambda_0 / 2 + \hbar^2 k^2 \beta / (4 m_e)  $, as well as a $k-$dependent band-shift term, 
$\mbb{E}_0^{\tau,s}(k) = \tau s \, \lambda_0 / 2 + \hbar^2 k^2 \alpha /(4 m_e) $, yielding 

\begin{equation}
\varepsilon^{\tau,s}_{\gamma}(k) = \mbb{E}_0^{\tau,s}(k) + \gamma \sqrt{
\left[ \Delta_0^{\tau,s}(k) \right]^2 + (t_0 a_0 k)^2}\, ,
\label{E1}
\end{equation}
where $\gamma = \pm 1$ determines the electron or hole state in complete analogy to graphene with or without a gap. 
By neglecting all the higher-order terms on order of ${\cal O}(k^4)$ for small $k$ values, Eq.\,(\ref{E1}) turns into

\begin{equation}
\varepsilon^{\tau,s}_{\gamma}(k)\backsimeq \frac{1}{2}\,\tau s\lambda_0 + \frac{\alpha\hbar^2}{4 m_e}\,k^2 + 
\frac{\gamma}{2}
\sqrt{
\left( \Delta - \tau s \, \lambda_0 \right)^2 + 
\left[
(2t_0a_0)^2+\left( \Delta - \tau s \, \lambda_0  \right)  \beta\hbar^2 / m_e
\right] k^2}\, .
\label{ksq}
\end{equation}
\medskip

Besides the simple plane-wave part, the spinor parts of the wave functions associated with the eigenvalues in Eq.\,(\ref{ksq}) for each valley are given by

\begin{equation}
\Psi^{\tau,s}_{\gamma}(k) =\frac{1}{\sqrt{2 \, \delta\varepsilon^{\tau,s}_{\gamma}(k)/ \gamma } } \left[
\begin{array}{c}
\sqrt{ \vert \delta\varepsilon^{\tau,s}_{\gamma}(k) + \Delta_0^{\tau,s}(k) \vert} \\
\\
\gamma\sqrt{ \vert \delta\varepsilon^{\tau,s}_{\gamma}(k)- \Delta_0^{\tau,s}(k) \vert} \, \tet{e}^{i\theta_k}
\end{array}
\right] \, ,
\label{wfunc}
\end{equation}
where $\theta_k =\tan^{-1}(k_y/k_x)$, $\delta\varepsilon^{\tau,s}_{\gamma}(k)\equiv\varepsilon^{\tau,s}_{\gamma}(k)-\mbb{E}_0^{\tau,s}(k)$,
and the overlap factor is calculated as

\begin{equation}
\mbb{F}^{\tau,s}_{\gamma,\gamma'}(k,\,k+q) \equiv \vert  \langle \Psi^{\tau,s}_{\gamma}(k) \, \vert \,
\Psi^{\tau,s}_{\gamma'}(\vert {\bf k} + {\bf q} \vert) \rangle \vert^2 =\frac{1}{2} \left[
1+ \gamma\,\gamma'\,\frac{\Delta_0^{\tau,s}(k)\Delta_0^{\tau,s}(\vert {\bf k} + {\bf q} \vert)  + { \bf k} \cdot ( {\bf k} + {\bf q})}{
\vert \delta\varepsilon^{\tau,s}_{\gamma}(k)\vert \, \vert \delta\varepsilon^{\tau,s}_{\gamma}(\vert {\bf k} + {\bf q} \vert) \vert}
\right]\, .
\label{overlap}
\end{equation}
Here, each part of the expression in Eq.\,(\ref{overlap}) could be calculated explicitly, e.g.,

\begin{equation}
\Delta_0^{\tau,s}(k) \, \Delta_0^{\tau,s}(\vert {\bf k} + {\bf q} \vert) = \frac{\left( \Delta - \tau s \, \lambda_0 \right)^2}{4} + 
\frac{\hbar^2 \beta(\Delta - \tau s \, \lambda_0)}{8 m_e} \, [2 { \bf k} \cdot ({\bf k} + {\bf q}) + q^2]
+\left(\frac{\hbar^2  \beta}{4 m_e}\right)^2 k^2 \,\vert{\bf k} + {\bf q} \vert^2 \, .
\end{equation}
\medskip

If we introduce the notation for a composite index $\mu\equiv\tau s=\pm 1$ and neglect the small $\alpha$ and $\beta$ terms in Eq.\,(\ref{ksq}), this gives rise to 
$\varepsilon^\mu_{\gamma}(k)\backsimeq \mu\lambda_0/2+\gamma\sqrt{(t_0a_0)^2k^2+(\Delta-\mu\lambda_0)^2/4}$, and therefore, the wave function in Eq.\,(\ref{wfunc}) could be simplified as 

\begin{equation}
\Psi^\mu_\gamma (k)=\frac{1}{\sqrt{[2\varepsilon^\mu_{\gamma}(k) - \mu\lambda_0] /\gamma}} \left[
\begin{array}{c}
\sqrt{\vert \varepsilon^\mu_{\gamma}(k)-\mu\lambda_0 + \Delta/2 \vert} \\
\\
\gamma\sqrt{\vert \varepsilon^\mu_{\gamma}(k)-\Delta/2  \vert} \, \tet{e}^{i\theta_k}
\end{array}
\right] =  \frac{1}{\sqrt{2\, \mbb{Q}_k}} \left[
\begin{array}{c}
\sqrt{\vert\mbb{Q}_k + \gamma(\Delta - \mu\lambda_0)/2\vert } \\
\\
\gamma\sqrt{\vert\mbb{Q}_k - \gamma( \Delta-\mu\lambda_0 )/2\vert } \, \tet{e}^{i\theta_k}
\end{array}
\right] \, ,
\label{wfunc-2}
\end{equation}
where $\mbb{Q}_k  = [\varepsilon^\mu_{\gamma}(k) - \mu\lambda_0/2]/\gamma = \sqrt{(t_0a_0)^2k^2+(\Delta-\mu\lambda_0)^2/4}$. It is clear from Eq.\,(\ref{wfunc-2}) that,
unlike gapped graphene, two spinor components become inequivalent and their ratio is different for electron and hole states. 
\medskip

Furthermore, the density of states $\rho_d(\mbb{E})$ can be formally written as

\begin{equation}
\rho_d(\mbb{E})= 2\int \frac{d^2{\bf k}}{(2 \pi)^2} \sum\limits_{\gamma = \pm 1}\,\sum\limits_{\mu= \pm 1} \, \delta\left[ \mbb{E} -
\varepsilon^\mu_{\gamma}(k) \right] \, .
\label{dos}
\end{equation}
By denoting $\breve{\epsilon}_\mu=\mu\lambda_0/2$, $\breve{\Delta}_\mu=(\Delta -\mu\lambda_0)/2$,
$\breve{A}_\mu=(\Delta -\mu\lambda_0)\hbar^2 \beta /(4 m_e) + (t_0a_0)^2$, and $\breve{\alpha}=\hbar^2\alpha/(4m_e)$, Eq.\,(\ref{ksq}) is further simplified to
$\varepsilon^\mu_{\gamma}(k)\backsimeq\breve{\epsilon}_\mu+\breve{\alpha}\,k^2 + \gamma\sqrt{\breve{\Delta}_\mu^2 + \breve{A}_\mu k^2}$.
Therefore, if $\breve{\alpha}\neq 0$, from Eq.\,(\ref{dos}) we obtain the analytical result for $\rho_d(\mbb{E})$, given by

\begin{equation}
\rho_d(\mbb{E}) = \frac{1}{2\pi} \, \sum\limits_{\pm} \, \sum\limits_{\gamma,\,\mu=\pm 1} 
\Big|
\breve{\alpha} + \frac{\gamma\breve{A}_\mu}{2[\mbb{E}-\breve{\epsilon}_\mu - \chi_\mu^\pm(\mbb{E})]} 
\Big|^{-1} 
\Theta \left[
\gamma\left(\mbb{E} -\frac{\mu\lambda_0}{2}\right) - \frac{1}{2} \left( \Delta - \mu\lambda_0 \right)
\right]\, ,
\label{dos-2}
\end{equation}
where $\Theta(x)$ is a unit-step function, and the energy-dependent function $\chi_\mu^{\pm}(\mbb{E})$ is defined as

\begin{equation}
\chi_\mu^{\pm}(\mbb{E}) = \frac{1}{2 \breve{\alpha}} \left[ \breve{A}_\mu + 2\breve{\alpha} (\mbb{E} - \breve{\epsilon}_\mu)  \pm
\sqrt{
\breve{A}_\mu^2+ 4 \breve{\alpha}^2 \breve{\Delta}_\mu^2 + 4 \breve{A}_\mu \breve{\alpha} (\mbb{E} - \breve{\epsilon}_\mu)}\,\right] \, .
\end{equation}
If $\breve{\alpha}=0$, on the other hand, we simply find

\begin{equation}
\rho_d(\mbb{E}) = \frac{2}{\pi} \, \sum\limits_{\gamma,\,\mu=\pm 1} 
\frac{|\mbb{E}-\mu\lambda_0/2|}{|\breve{A}_\mu|} 
\Theta \left[
\gamma\left(\mbb{E} -\frac{\mu\lambda_0}{2}\right) - \frac{1}{2} \left( \Delta - \mu\lambda_0 \right)
\right]\, .
\label{dos-3}
\end{equation}
Using the result in Eq.\,(\ref{dos-2}), all previously known cases, including a pair of parabolic bands or Dirac cones, as well as a pair of gapped Dirac cones, could be easily verified. 
\medskip

If we neglect the $\alpha$ and $\beta$ terms in Eq.\,(\ref{ksq}), i.e., setting $\breve{A}_\mu=(t_0a_0)^2$ and $\breve{\alpha}=0$, we obtain from Eq.\,(\ref{dos-3})
the result for a pair of non-degenerate, spin- and valley-dependent subbands in gapped graphene, given by

\begin{equation}
\rho_d(\mbb{E}) = \frac{2}{\pi(t_0a_0)^2} \sum \limits_{\gamma,\,\mu = \pm 1} \Big| \mbb{E} -\frac{\mu\lambda_0}{2}  \Big| \,\,\, 
 \Theta \left[
 \gamma\left(\mbb{E} -\frac{\mu\lambda_0}{2}\right) - \frac{1}{2} \left( \Delta - \mu\lambda_0 \right)
 \right]\, .
 \label{dos-4}
\end{equation}
It is clear from Eqs.\,(\ref{dos-2}), (\ref{dos-3}) and (\ref{dos-4}) that the boundaries for non-zero density of states in all three cases are set by
$\mbb{E} > \Delta/2$ for electrons ($\gamma = +1$) and $\mbb{E} < - \Delta / 2 + \mu\lambda_0 $ for holes ($\gamma = - 1$). 
\medskip

For weak hopping with $t_0\ll\Delta$, using Eq.\,(\ref{ksq}) we arrive at 
 
\begin{equation}
\varepsilon_{\gamma}^{\mu}(k)  = \frac{1}{2} \left[ 
\mu\lambda_0 (1 - \gamma) + \gamma \, \Delta 
\right] + \left[
\frac{\hbar^2}{4 m_e} (\alpha + \gamma \beta ) + \frac{\gamma \, (t_0a_0)^2}{\Delta -\mu\lambda_0}
\right] k^2 \, ,
\label{parab}
\end{equation}
where we have used the fact that $\Delta\gg\lambda_0$. This result leads to the density of states given by

\begin{equation}
\rho_d(\mbb{E}) = \frac{1}{2\pi\hbar^2} \sum \limits_{\gamma,\,\mu = \pm 1}
\Big| \frac{\alpha+\gamma \beta}{4 m_e} + \frac{\gamma (t_0a_0)^2}{ \hbar^2(\Delta - \mu\lambda_0)} \Big|^{-1}
\Theta \left[
\gamma\left(\mbb{E} -\frac{\mu\lambda_0}{2} \right) - \frac{1}{2} \left( \Delta -\mu\lambda_0 \right)
\right]\, ,
\end{equation}
where there exist two energy-independent giant discontinuities for electrons and holes, respectively.  
\medskip

\begin{figure}
\centering
\includegraphics[width=0.49\textwidth]{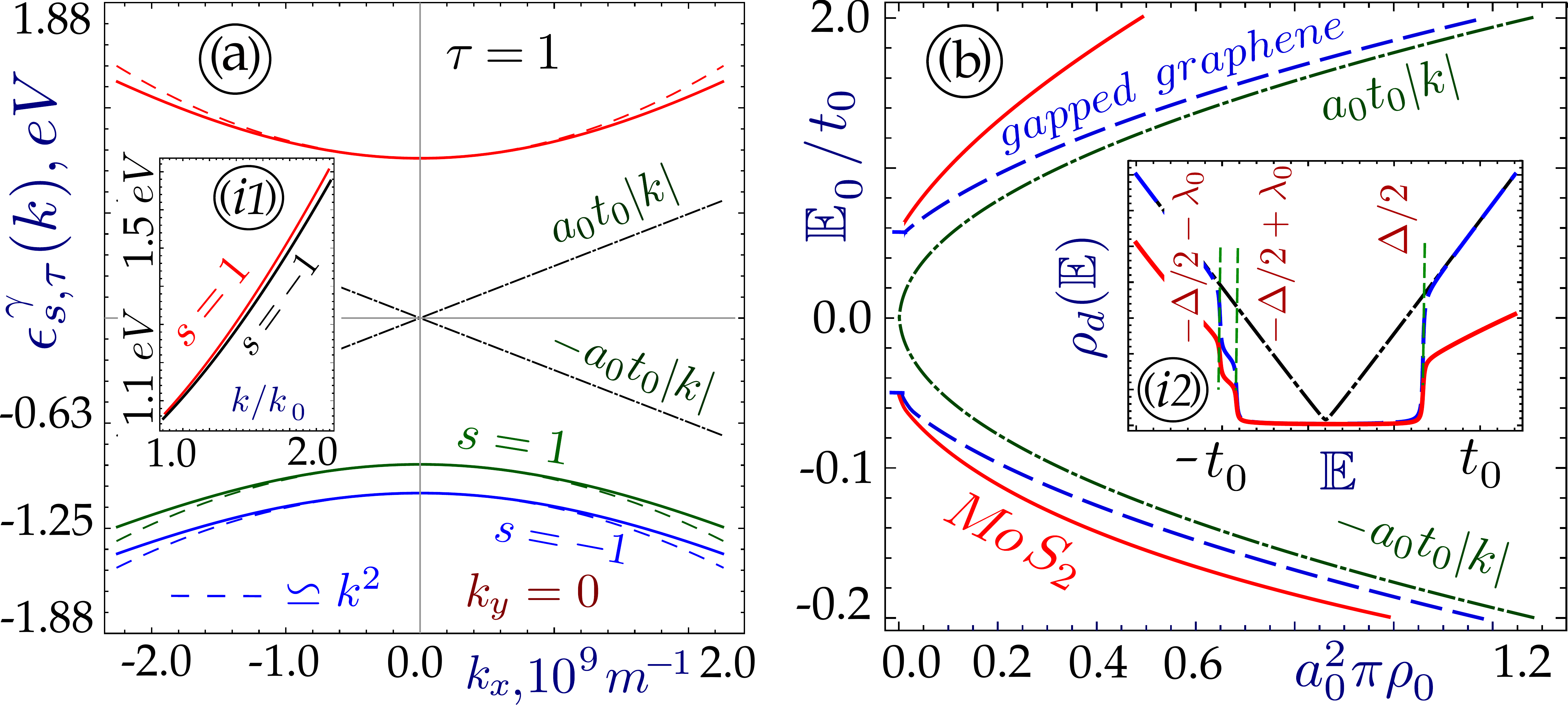}
\caption{Electron and hole energy dispersions (solid curves) and their Fermi energies for a monolayer MoS$_2$. Plot ($a$) represents the dispersions of energy subbands near
the $K$-valley ($\tau = 1$), where the corresponding results in the parabolic approximation are also shown by dashed curves. In addition, the Dirac-cone dispersions $\varepsilon_\gamma(k) = \gamma\,t_0 a_0 \vert k \vert$ 
are included. The inset $(i1)$ gives a close-look view for two very close conduction bands with opposite spins $s=\pm 1$. 
Plot ($b$) presents the calculated Fermi energies $E_F/t_0$ for electrons ($\gamma=+1$) [holes ($\gamma=-1$)] 
as a function of electron (hole) doping density $a_0^2\pi\rho_0$ in MoS$_2$ (red solid curve with neglected $k^2$-terms in the Hamiltonian), gapped graphene (blue dashed curve) 
and Dirac cones (green dash-dotted curve) outside the gap region. 
The corresponding density-of-states curves for these materials are displayed in the inset $(i2)$.}
\label{FIG:7}
\end{figure}

For electrons with $\gamma=+1$ at $\mbb{E}= \Delta/2$, we get the jump in the density of states given by

\begin{equation}
\delta\rho_d^{\gamma =+1} = \frac{1}{2 \pi} \sum\limits_{\mu= \pm 1}
\left[  
\frac{(t_0a_0)^2}{\Delta -\mu\lambda_0} +
\frac{(\alpha + \beta) \hbar^2}{4 m_e} 
\right]^{-1} = \frac{0.18}{t_0a_0^2}\, .
\end{equation}
Similarly, for holes with $\gamma=-1$ at $\mbb{E}= -\Delta/2+\mu\lambda_0$, we obtain two discontinuities at different energies, i.e.,

\begin{equation}
\delta\rho^{\gamma=-1}_d(-\Delta/2+\mu\lambda_0) = 
\frac{1}{2\pi\hbar^2}\left[\frac{\alpha-\beta}{4m_e}-\frac{(t_0a_0)^2}{\hbar^2}\left\{
\begin{array}{ll}
\times (\Delta-\lambda_0)^{-1}\\
\\
\times (\Delta+\lambda_0)^{-1}
\end{array}\right.\right]^{-1}
=\frac{1}{t_0a_0^2}\left\{\begin{array}{ll}
\times 0.104\, , &  \mbox{for $\mu=+1$}\\
\\
\times 0.110\, , &  \mbox{for $\mu=-1$}
\end{array}\right.\, .
\end{equation}
Our above analytical expressions match exactly
the numerical results reported in Ref.\,[\onlinecite{mos2}]. We note that the simplified parabolic dispersions in Eq.\,(\ref{parab}) catch 
the difference in electron and hole effective masses for various spin and valley indices even for $k 
\Longrightarrow 0$. For large $k$ values, this difference becomes more significant. The numerically-calculated energy dispersions for electrons and holes and their zero-temperature Fermi energies for 
fixed doping density are presented in Fig.\,\ref{FIG:7}. It is interesting to note that at a finite energy away from the bandedge, the density of states 
for $\mathrm{MoS}_2$ is significantly smaller compared to graphene. This agrees with the well-known fact that non-parabolicity
in graphene energy subbands will enhance the density of states. 
\medskip

Now, let us turn to studies of nonlocal plasmon dispersions and their damping in a $\mathrm{MoS}_2$ layer interacting with a semi-infinite conductor. For this case, 
the one-loop electron polarization function for molybdenum disulfide with two pairs of energy subbands can be obtained in a way similar to Eq.\,(\ref{sum})
by summing over a composite index $\mu$ for each subband but specifying $\gamma$ values for $n-$ and $p-$doping separately.
We will further assume that only the lowest hole subband with $\mu=+1$ will be occupied, and two electron subbands become nearly degenerate with each other since $\Delta\gg\lambda_0$.
\medskip

In the long-wavelength limit, by using Eq.\,(\ref{longw}) the polarization function of a $\mathrm{MoS}_2$ monolayer interacting with a semi-infinite 
conductor separated by a distance $\mathcal{D}$ can be expressed as 

\begin{eqnarray}
\nonumber
&& \omega^\gamma_{p,-}(q) = q\,\sqrt{2 \mathcal{D} \, \mbb{L}(\gamma)} \backsimeq q\, \sqrt{\pi\alpha_0 \rho_0 \mathcal{D} \, (\gamma+3)} \,  \frac{2(t_0 a_0)^{3/2}}{\hbar\sqrt{\Delta - 
(1 - \gamma) \lambda_0}}\, ,\\
&& \omega^\gamma_{p,+}(q) = \frac{\Omega_p}{\sqrt{2}} + \frac{\mbb{L}(\gamma)\,q}{\sqrt{2}\, \Omega_p} \backsimeq \frac{\Omega_p}{\sqrt{2}} + 
\frac{\gamma+3}{\sqrt{2}\, \Omega_p}\, (t_0a_0)^3 \frac{2\pi \alpha_0 \rho_0}{\hbar^2\left[\Delta - (1 - \gamma) \lambda_0 \right]} \,q \, ,
\label{mopl}
\end{eqnarray}
where $\rho_0$ is the areal density for doping,
$\alpha_0 = e^2/(4\pi\epsilon_0\epsilon_rt_0a_0)\backsimeq 4.9$ is the fine-structure constant,
$\epsilon_r\backsimeq 5$ is the dielectric constant for $\mathrm{MoS}_2$, and

\begin{equation}
\mbb{L}(\gamma)=\frac{2\pi(t_0a_0)^3\alpha_0 \rho_0(\gamma+3)}{\hbar^2[\Delta+(1-\gamma)\lambda_0]}\, .
\end{equation}
Here, the inclusion of the coupling between $\mathrm{MoS}_2$ and the semi-infinite conductor has split plasmons into in-phase ($+$) and out-of-phase ($-$) modes in Eq.\,(\ref{mopl}).  
This will certainly lead to a modification of plasmon-mode damping by PHMs.
\medskip

The main advantage for using $\mathrm{MoS}_2$ in a hybrid plasmonic device is its large energy gap $\backsimeq \Delta$, which allows one
to consider clean metals with an extremely high plasma frequency $\hbar \Omega_p \backsimeq 1\,eV$. This arrangement is not 
possible for gapped graphene or silicene since
the plasmon modes at such a frequency would be strongly damped by the interband PHMs. Another unique feature for $\mathrm{MoS}_2$ is the large difference between the 
electron and hole doping processes, i.e., high doping density $\rho_0 \backsimeq 10^{11}-10^{13}\,$cm$^{-2}$ only allows the occupation of one hole subband, as assumed in Eq.\eqref{mopl} for $\gamma=-1$. 
Here, even in the parabolic approximation, the results for $n-$ and $p-$doping still vary drastically. Although the $\lambda_0$ 
correction to $\Delta$ is very small, the density of states of electrons is almost twice as large as that of holes. 
\medskip

In order to determine Landau damping of the plasmon modes, we need to determine the boundaries $\hbar\Omega_c^\gamma(q)$ for PHMs, defined by 

\begin{equation}
\hbar \omega^\gamma_{p,\pm}(q) \geq \hbar\Omega_c^\gamma(q)\equiv\varepsilon_\gamma^\mu(k_F + q)-\varepsilon_\gamma^\mu(k_F) \, ,
\end{equation}
which corresponds to ${\bf k}\pa{\bf q}$. Here, $k_F$ is the electron Fermi wave number.
For moderate $n-$doping ($\gamma=+1$), from Eq.\,(\ref{ksq}) its PHM boundary is found to be

\begin{equation}
\hbar \Omega^{\gamma=+1}_c(q) =  
\frac{\lambda_0}{2} + \frac{\hbar^2\alpha}{4 m_e} \left(q + k_F\right)^2 -\mbb{E}_F+ 
\frac{1}{2}
\sqrt{
\Delta^2 + 
\left(
\frac{\hbar^2 \beta \Delta}{m_e} + 4  t_0^2 a_0^2\right) 
\left(q + k_F\right)^2}\, .
\label{bound}
\end{equation}
In the long-wavelength limit with $q \ll k_F$,  we can approximate Eq.\,(\ref{bound}) by

\begin{equation}
\hbar\Omega^{\gamma=+1}_c(q)\backsimeq q\,
\sqrt{
 \left[4\,t_0^2 a_0^2+\frac{\hbar^2\Delta}{m_e}(\beta+\alpha)\right]\frac{\left(E_F-\Delta/2\right)}{\Delta}}\, ,
\end{equation}
where we use the facts that $\Delta\gg\lambda_0,\,t_0a_0k_F$ and $\hbar^2\beta k_F^2/m_e$ and $E_F$ is determined by $\rho_0$.
\medskip

Alternatively, if the sample is $p-$doped ($\gamma=-1$), the PHM boundaries $\hbar\Omega_c^{\gamma=-1}(q)$ with $\mu=+1$ for the occupied hole subband is found to be

\begin{equation}
\hbar\Omega_c^{\gamma = -1}(q) \backsimeq q\,
\sqrt{
\left[4\,t_0^2 a_0^2+\frac{\hbar^2(\Delta-\lambda_0)}{m_e}(\beta-\alpha)\right]\frac{\left(E_F-\Delta/2+\lambda_0\right)}{(\Delta-\lambda_0)}}\, ,
\end{equation}
These two PHM boundaries, $\hbar\Omega_c^{\gamma =\pm 1}(q)$, determine whether the acoustic-like plasmon branch would be Landau damped or not. On the other hand, the optical-like plasmon
branch originating from $\hbar\Omega_p/\sqrt{2}$ is considered to be far away from the interband PHM boundary starting around $\Delta \backsimeq 1.9\,$eV.

\section{Summary and concluding remarks}
\label{sect4}
 
In conclusion, we have presented in this paper the numerical results for full ranges of hybrid plasmon-mode dispersions, 
as well as analytical expressions in the long-wavelength limit, in an open interacting system including a 2D material and a conducting substrate. 
Although the plasmon damping is set by the particle-hole modes(PHMs) of electrons in the 2D material, the strong coupling between electrons in 2D materials and in the conducting substrate 
gives rise to a splitting of plasmons into one in-phase and one out-of-phase mode. Such dramatic changes in plasmon dispersions are expected to have impacts on the damping of these modes.    
In addition, in comparison with gapped graphene, the different plasmon modes in silicene or transition-metal dichalcogenides make our damping studies even more distinctive, 
including different energy bandgaps, doping types, occupations of subbands, and coupling between 2D materials and the conducting substrate. 
Here, each plasmon branch and its damping can be independently analyzed based on the signatures of the PHMs since the plasmon modes depend on both spin and valley degrees of freedom.  
Therefore, our proposed hybrid systems in this paper are expected to be useful in measuring the dielectric property of 2D material open systems (2DMOS) and spin-orbit coupling strength of individual 2D materials. 
More importantly, we have demonstrated the possibility to design the plasmonic resonances at almost all frequencies and wave numbers for different types of newly discovered 2D materials. This was not feasible for either
a free-standing silicene layer or a graphene-based hybrid structure.
\medskip

Additionally, our model and numerical results for 2DMOS have confirmed a recently discovered phenomenon related to a significant damping of an acoustic-like plasmon branch as the separation to the conducting substrate 
becomes very small. From our current studies, we have found that in silicene this critical distance can be modified by either applying an external electric field or varying doping types and levels.
The unique linear dispersion obtained under the long-wavelength limit makes the damping from intraband PHMs possible in 2DMOS but not for a free-standing 2D layer.  
We have also noted that the plasma energies in clean metals are usually much larger than the  Fermi energies and bandgaps in graphene. As a result, the plasmon modes in graphene can not be coupled 
to surface plasmons in the presence of a metallic substrate without suffering from the strong damping by interband PHMs. 
However, the use of $\mathrm{MoS}_2$ with a large bandgap in 2DMOS is able to suppress this damping effectively. 
Alternatively, one could also use Bi$_2$Se$_3$ material,\,\cite{pol01} which is a doped topological insulator with a surface-plasmon energy around $104\,$meV, or a highly-doped semiconductor in 2DMOS.

\acknowledgments

D.H. would like to thank the support from the Air Force 
Office of Scientific Research (AFOSR).
We would like to mention a great help from Joseph Sadler and especially Patrick Helles, IT managers at the Center for High Technology Materials of the University of New Mexico, much beyond their 
official responsibilities.

\bibliography{Slb}

\end{document}